 \documentclass[preprint2]{aastex}

\newcommand{\bgc}{$B_{gc}$}
\newcommand{\Tx}{$T_{\rm X}$}
\newcommand{\lx}{$L_{\rm X}$}
\newcommand{\Lx}{$L_{\rm X}$}
\newcommand{\tx}{$T_{\rm X}$}
\newcommand{\gtapr} {\lower .1ex\hbox{\rlap{\raise .6ex\hbox{\hskip .3ex
        {\ifmmode{\scriptscriptstyle >}\else
                {$\scriptscriptstyle >$}\fi}}}
        \kern -.4ex{\ifmmode{\scriptscriptstyle \sim}\else
                {$\scriptscriptstyle\sim$}\fi}}}
\newcommand{\ltapr} {\lower .1ex\hbox{\rlap{\raise .6ex\hbox{\hskip .3ex
        {\ifmmode{\scriptscriptstyle <}\else    
                {$\scriptscriptstyle <$}\fi}}}
        \kern -.4ex{\ifmmode{\scriptscriptstyle \sim}\else
                {$\scriptscriptstyle\sim$}\fi}}}
\begin{document}
\title{Correlations of Richness and Global Properties in
Galaxy Clusters}
\author{H.K.C. Yee\altaffilmark{1}}
\affil{Department of Astronomy and Astrophysics, University 
of Toronto, Toronto, ON, M5S 3H8, Canada.}
\email{hyee@astro.utoronto.ca} 
\author{E. Ellingson\altaffilmark{1}}
\affil{CASA, University 
of Colorado, Campus Box 389, Boulder, CO 80309}
\email{elling@casa.colorado.edu} 
\altaffiltext{1}{Visiting Astronomer, Canada-France-Hawaii Telescope,
which is operated by the National Research Council of Canada,
Le Centre National de Recherche Scientifique, and the University
of Hawaii.}

\received{}
\accepted{}

\begin{abstract}
Richness is a key defining characteristic of a galaxy cluster.
We measure the optical richness of galaxy clusters from the CNOC1 cluster
redshift survey using the galaxy-cluster center correlation
amplitude \bgc.
We show that the \bgc~values measured using photometric
catalogs are consistent with those derived from redshift
catalogs, indicating that richness can be measured reliably from
photometric data alone, even at moderate redshifts of $\sim$0.6.
We establish the correlations between optical richness and other important
attributes of a galaxy cluster, such as velocity dispersion, mass,
radius, and X-ray temperature and luminosity.
We find that the scaling relations of these quantities with richness 
are entirely consistent with those derived by assuming 
a simple mass density profile at 0.5 $h_{50}^{-1}$ Mpc of $\rho\sim r^{-1.8}$.
The excellent correlations between \bgc~and velocity dispersion
and X-ray temperature allow one to use richness, an easily measurable
quantity using relatively shallow optical imaging data alone, 
as a predictor of these quantities at moderate redshifts.
The \bgc~parameter can be used to estimate the velocity dispersion
of a cluster to a precision of approximately 15\% 
($\sim\pm100$ km s$^{-1}$), and X-ray temperature to about 20\%.
Similar correlations, but with larger scatter, are also obtained
between richness and the characteristic radius and mass of the clusters.
We compare the relative merits of \bgc, \tx, and \lx~as predictors of the
dynamical mass, and find that they are comparable, providing 
estimates at an accuracy of $\sim30$\%.
We also perform similar analyses of correlations between richness
and velocity dispersion, \tx, and \lx~with a sample of low-redshift
Abell clusters and find consistent results, but with larger
scatter, which may be the result of a less homogeneous 
database, or sample-dependent effects.

\end{abstract}

\section{Introduction}

The single most important attribute of a galaxy cluster is its mass.
The cluster mass function and its evolution provide constraints to
the evolution of large scale structure and important cosmological parameters
such as $\Omega_m$ and $\sigma_8$ (e.g., Carlberg et al.~1997; Oukir \&
Blanchard 1997; Fan
et al.~1997; Viana \& Liddle 1999; Borgani et al.~2000; and many others),
and possibly $w_\phi$, the equation of state of dark matter  
(e.g., Levine, Shultz, \& White 2002).
The mass-to-light ratio of clusters provides one of the most robust dynamical
determinations of $\Omega_m$ via the Oort (1958) method (e.g., 
Carlberg et al.~1996; Girardi et al.~2000).
Therefore, it is not surprising that over the last 70 years, 
starting with Zwicky (1937), much effort has been spent measuring the mass
of clusters using a number of techniques. These include dynamical methods
using redshift surveys (e.g., Kent \& Gunn 1982; Carlberg et al. 1996; 
Girardi et al.~1998, Rines et al. 2000), measuring the temperature of 
hot intracluster gas
using X-ray observations (e.g., Allen 1998; Lewis et al.~1999),
weak gravitational lensing (e.g., Hoekstra et al.~1998; Clowe et al.~2000),
and observations of the Sunyaev-Zeldovich effect (e.g., Grego
et al.~2001).
These methods in general are expensive in terms of the
 observational resources required,
especially at higher redshifts.

There have been a number of comparisons made between the
different methods for determining the mass of clusters.
These include comparisons between
the X-ray and strong lensing methods in cores of clusters
 (e.g., Allen 1998; Wu 2001); between the X-ray and weak lensing
methods (e.g., Smail et al.~1997); and between
the dynamical and X-ray methods (e.g., Girardi et al.~1998).
Of particular interest is the
detailed comparison of the first three of
these methods using the CNOC1 (Canadian Network for Observational Cosmology)
sample of clusters performed by Lewis et al.~(1999).
Their results demonstrated that consistent mass estimates
outside the cluster core region
can be obtained by the dynamical, X-ray, and weak lensing methods.

Naively, in the situation where light traces mass,
the richness of a galaxy cluster should be a direct indication
of its mass.  Furthermore, observationally speaking,
richness should be a relatively inexpensive parameter
to measure, requiring only direct images of moderate depth even for
$z\sim1$ clusters.
Historically, however, richness has not been a well-defined,
quantitatively measured parameter.
Yee \& L\'opez-Cruz (1999, here after YL99) examined in detail the properties
of the richness parameter \bgc, first used by Longair \& Seldner
(1979) for quantifying the environment of radio galaxies.
\bgc~is the amplitude of the galaxy-cluster center correlation function 
measured individually for each cluster, and basically scales as the
net counts of galaxies normalized by the luminosity function and
spatial distribution of the cluster galaxies (see Seldner \& Longair
and YL99).
Yee \& L\'opez-Cruz (LY99), using photometric catalogs of a large sample
of low-redshift Abell clusters, showed that with reasonably careful 
attention to the choice of the cluster galaxy luminosity 
function (LF) and background corrections, 
the \bgc~parameter is a robust measure of
the richness,  producing systematics of only 10 to 20\%~arising from the typical
systematic uncertainties in the LF, the photometry, and the spatial 
distribution of the cluster galaxies.

In this paper, we investigate the correlation of the optical richness 
of clusters with important cluster attributes such as mass, line-of-sight
velocity dispersion ($\sigma_1$),  X-ray temperature (\Tx) 
and luminosity ($L_{\rm X}$), and virial radius ($r_{vir}$),
using primarily the CNOC1 Cluster Redshift Survey sample (Yee, 
Ellingson, \& Carlberg 1996),  currently the largest
set of homogeneous spectroscopic and photometric data
of galaxy clusters at the moderate redshift range of 0.17 to 0.55.
We show that a well-defined and easily measurable
 richness parameter such as \bgc~produces
strong correlations with other global properties of galaxy clusters,
allowing its use as an estimator for
quantities such as velocity dispersion, mass, \Tx, and \lx.
We demonstrate that \bgc~and the more conventional
\tx~and \lx~are good predictors of the
virial mass of the clusters at the $\sim$30\% level, 
with \bgc~and \tx~being somewhat superior to \lx.
The correlations of the optical richness parameter with these key 
properties make richness an important
defining parameter for galaxy clusters.
As we enter into an era of wide-field, deep optical
imaging capability, more effective and efficient methods of finding
galaxy clusters using optical techniques are being investigated and
applied (e.g., Gladders \& Yee 2000; Kim et al.~2002).
Optical cluster surveys covering from tens to thousands of square 
degrees with sample size as large as $10^4$ clusters are  becoming 
available (e.g., the Red-Sequence Cluster Survey, see Yee \& Gladders 2001;
the SDSS, see Bahcall et al.~2002).
The ability to use optical data alone, via richness, to estimate
important quantities such as mass, velocity dispersion, and X-ray
temperature and luminosity for clusters at large
redshifts, becomes increasingly important in 
the characterization of cluster samples, as 
obtaining X-ray or comprehensive spectroscopic data becomes impractical
for such a large number of clusters.

The remainder of the paper is organized as follows.
In \S2, we briefly describe the optical and X-ray data sets.
Section 3 presents the measurements of the richness parameter \bgc~for the
clusters, including a comparison of the values determined using
photometric and spectroscopic data.
Section 4 examines the correlations between the richness of clusters
with various global properties.
We discuss the implications and applications of these findings in \S5, and
summarize our conclusions in \S6.
Throughout this paper, unless otherwise noted,
we assume a $\Lambda=0$ cosmology with $q_0=0.1$,
 and express $H_0$ in units of $h_{50}=50$ km s$^{-1}$ Mpc$^{-1}$.

\section{Cluster Samples and Data} 

The CNOC1 Cluster Redshift Survey is a multi-object 
spectroscopy survey conducted at the
Canada-France-Hawaii  3.6m telescope (CFHT) of galaxies 
in moderately rich to rich clusters
with a redshift range of 0.17 to 0.55.
The sample consists of 15 high X-ray luminosity Einstein Medium Sensitivity 
Survey (EMSS) clusters (Gioia et al.~1990) plus Abell 2390.
This sample has currently the most comprehensive and homogeneous
optical and X-ray database, making it an excellent sample for
the investigation of the correlations between cluster richness and
other global properties.

The  observational strategy and procedure and the data
reduction techniques for the photometric and redshift catalogs
of cluster galaxies are described in detail in Yee et al.~(1996).
The complete catalogs contain about 2600 redshifts, of which about
half are considered cluster members.
Examples of data catalogs of the survey can be found in Yee et al. (1998),
Ellingson et al.~(1998), Abraham et al.~(1998) and others.
The velocities have a typical accuracy of $\sim$135 km s $^{-1}$.
Photometry in Gunn $r$ and $g$ is also available for all galaxies.
An important feature of the CNOC1 redshift survey is the estimate of
statistical weights for the redshift catalog objects.
The redshift survey used a sparse sampling strategy of about 1 in 2.
The establishment of statistical weights of each galaxy as
a function of magnitude, position, and color, allows the use of the
redshift sample as a complete sample.
Velocity dispersions and dynamically determined masses for the clusters are
presented in Carlberg et al.~(1996).
The cluster MS0906+11, shown to be a binary in velocity space,
(making proper dynamical analysis difficult), is
excluded from the dynamical sample, leaving a total of 15 clusters (Table 1).

X-ray temperatures for the CNOC1 clusters were taken from the
literature, primarily from Mushotzky \& Scharf (1997) and
Henry (2000).
Sources for other temperature measurements are listed in
Lewis et al.~(1999). These temperatures are predominantly
emission-weighted measures from the Advanced Satellite for Cosmology and 
Astrophysics (ASCA), without correction for
non-isothermality in the cluster cores.

X-ray luminosities for these clusters are estimated
from those calculated by Ellis \& Jones (2002) with an additonal
correction for cooling flows.
The Ellis \& Jones luminosities
include an additional aperture correction over
the original luminosity estimates from Gioia \& Luppino (1994),
particularly for clusters with large core radii. For clusters
which are not listed in Ellis \& Jones, we use a similar
correction to estimate aperture-corrected EMSS luminosities.
For this correction, core radii were taken from Lewis et al.~(1999) from ROSAT
High Resolution Imager (HRI) observations to determine the spatial profiles 
of the X-ray emission. These estimates should be generally consistent 
with those from Ellis \& Jones. The luminosity of Abell 2390 was estimated 
from data from Pierre et al. (1996) and includes a similar aperture correction.
We correct these EMSS luminosities to bolometric
luminosities using $N_H$ values from Dickey \& Lockman (1990) and
temperatures derived from ASCA and Chandra observations from the
literature. 
An X-ray temperature is not available for MS1231+15; we assume a typical
value of 6 keV for this correction.

Finally, we apply a rudimentary correction for the effects of
core cooling flows, based on the HRI spatial
profiles from Lewis et al.~(1999). For this correction, the central regions 
of clusters with cooling flows were omitted from beta-model fits to the 
outer profiles and the excess flux over the fit was attributed to the
cooling flow. These corrections may be a slight overestimate of the effects 
of the cooling flow, however, as the spatial profiles were performed in
the slightly softer ROSAT energy band. However, these corrections are on
the order of 2-19 percent of the total flux, and so represent a small
additional uncertainty.
The X-ray temperatures and luminosities are listed in Table 1.

A significant amount of data exists for lower redshift Abell clusters
which can be used for comparison with the CNOC1 results.
In particular, the LOCOS (Low-redshift Cluster Optical Survey) sample
of L\'opez-Cruz \& Yee (see L\'opez-Cruz 1997 and YL99) 
provides a large homogeneous optical database
for 46 Abell Clusters at $z\sim0.05$.
The photometric data, based on KPNO 0.9m CCD images in $I_c$, $R_c$, and $B$,
were produced using the same procedure as that of the CNOC1 survey.
We have culled from the literature data on velocity dispersion, \Tx, and
\lx~for clusters in this sample to allow us to compare the correlations
of richness with global properties for the
moderate redshift CNOC1 sample with a zero redshift sample.

\begin{deluxetable}{lcccccccc} 
\tablenum{1} 
\tablewidth{6.5true in} 
\tablecaption{CNOC1 Clusters} 
\tablehead{ 
\colhead{Cluster} & 
\colhead{$z$} & 
\colhead{$B_{gc}$\tablenotemark{a}\tablenotetext{a}{
\bgc~in units of $h_{50}^{-1.8}$ Mpc$^{1.8}$.}
} & 
\colhead{$\pm$}   &
\colhead{Abell Class\tablenotemark{b}\tablenotetext{b}{
Abell richness class estimated from \bgc.}
}  &
\colhead{$T_{\rm X}$ (keV)} &
\colhead{$\pm$}   &
\colhead{$L_{{\rm X}bol}$\tablenotemark{c}\tablenotetext{c}{
Bolometric X-ray luminosity in units of 10$^{44}$ $h_{50}^{-2}$ erg s$^{-1}$.}
} &
\colhead{$\pm$}   
} 
\startdata 
 MS0016+16 &    0.5465 &   2100 &  305 & 4 & 8.0 & 0.3 & 48.00 & 1.22 \\
 MS0302+16 &    0.4245 &   ~782 &  206 & 0 & 4.4 & 0.7 & ~6.60 & 1.56\\
 MS0440+02 &    0.1965 &   ~616 &  170 & 0 & 5.3 & 0.5 & ~7.10 & 1.50 \\
 MS0451+02 &    0.2011 &   1275 &  222 & 2 & 8.6 & 0.3 & 15.47 & 3.30 \\
 MS0451--03 &   0.5391 &   2060 &  301 & 4 & 10.4 & 0.75 & 55.40 &  2.64\\
 MS0839+29 &    0.1930 &   1318 &  223 & 2 & 4.2 & 0.2 & ~7.87 & 1.49 \\
 MS1006+12 &    0.2604 &   1474 &  239 & 2 & 8.5 & 1.0 & 13.26 & 3.10  \\
 MS1008--12 &   0.3063 &   1757 &  258 & 3 & 7.3 & 1.0 & 14.97  & 3.40 \\
 MS1224+20 &    0.3255 &   ~508 &  176 & 0 & 4.3 & 0.65 & ~5.86  & 1.34 \\
 MS1231+15 &    0.2353 &   ~904 &  196 & 1 & --- & ---  & 11.85 & 2.68 \\
 MS1358+62 &    0.3290 &   1373 &  238 & 2 & 6.5 & 0.3  & 14.85 & 0.66 \\
 MS1455+22 &    0.2568 &   ~593 &  176 & 0 & 5.5 & 0.15 & 24.28  & 2.02 \\
 MS1512+36 &    0.3717 &   ~610 &  189 & 0 & 3.6 & 0.6  & ~6.17  & 1.30 \\
 MS1621+26 &    0.4275 &   1158 &  231 & 1 & 6.6 & 0.9  & 18.90 & 1.85 \\
 A2390     &    0.2280 &   1604 &  243 & 3 & 8.9 & 0.35 & 45.43 & 14.0 \\
\enddata
\end{deluxetable} 

\section {Richness Measurements of the Clusters} 

We use the parameter \bgc~(see Longair \& Seldner 1978; 
Andersen \& Owen 1996; YL99)
as an estimate for the richness of the clusters.
The \bgc~parameter, defined as the galaxy-cluster center correlation
amplitude (i.e., $\xi(r)=B_{gc}r^{-\gamma}$), essentially measures
the net number of galaxies in a cluster within some fixed aperture,
scaled by a luminosity function and a spatial distribution function.

Yee \& L\'opez-Cruz (YL99) have shown using a
large sample of low-redshift Abell clusters that 
\bgc~is a robust richness estimate, given that the photometry is
accurate to $\sim 0.2$ mag and the LF of cluster
galaxies is known and universal to within about 20\% accuracy in
the Schechter function parameters $M^*$ and $\alpha$.
Furthermore, they also showed that \bgc~values computed using different
absolute magnitude limits (but at least approximately a magnitude past $M^*$)
and different sampling radii are stable to within the uncertainty.
Yee \& L\'opez-Cruz also
provided a calibration of the \bgc~value to the more traditional
Abell Richness Class.
If the evolution of the cluster galaxy LF is known, \bgc~effectively
allows the uniform computation of cluster richness for different
redshifts, providing a powerful parameter for the comparison of cluster
samples spanning a significant redshift range.

We measure the \bgc~parameter for the CNOC1 clusters via two
methods: using the photometric catalogs and the redshift catalogs.
This allows us to verify the accuracy of the photometric method
which uses a statistical background count correction.

\subsection{Photometric \bgc}

In the standard photometric method (see, e.g., YL99)
we count galaxies to a fixed absolute magnitude within
a standard cluster-centric radius and correct
the expected background counts to obtain the net counts.
We follow the prescription of YL99 for k-correction, and
count galaxies to a k- and evolution-corrected $M_r$ of --20.0
over a radius of 0.5 $h_{50}^{-1}$ Mpc (physical).
We note that by counting to a relatively small radius, 
uncertainties arising from projection
and stochastic background variations can be kept to a minimum.
For background correction, we use counts in the $r$ band
generated from images of five randomly pointed fields observed
during the CNOC1 survey.
The absolute magnitude limit is chosen to be about 2 magnitudes
below $M^*$, which is a relatively slowly varying part of the LF,
providing the most stable result (see YL99).

The position of the brightest cluster galaxy (BCG) is chosen as the center
of the cluster in the computation of the \bgc~parameter (with the BCG
itself not included as part of the excess counts).
We note that in most cases the BCG can be determined unambiguously 
without redshift membership information by choosing the 
brightest galaxy on the red-sequence of the early-type galaxies of
the cluster on a color-magnitude diagram.
Carlberg et al.~(1999) show that for the CNOC1 clusters, the choice of
using the BCG or galaxy density centroid as the center of the cluster
produces minimal effect on the determination of the profile of the
galaxy cluster.
The use of the BCG as the center for the computation of \bgc~provides
the advantage of simple consistency that can be easily applied.

An important input in the estimate of \bgc~is the cluster galaxy LF
and its evolution over the redshift range of the sample.
Ideally, the most robust result is obtained when the LF (and
its evolution) is determined directly from the cluster sample itself.
However, despite being the currently largest cluster galaxy redshift sample
at these intermediate redshifts, the CNOC1 sample is not sufficiently
large at the higher redshift end to determine a statistically
significant LF evolution measurement.
Thus, we use the estimated CNOC2 overall field galaxy luminosity evolution
(Lin et al.~1999) with
$M^*(z)\sim M(0)+Qz$, where we adopt $Q\sim1.4$ for the cluster galaxy LF.
The zero redshift LF is taken to be a Schechter Function of 
$M^*_r(0)=-$21.9, and $\alpha=-1$, which is the LF used by
YL99 transformed to the Gunn system.
We note that because \bgc~is derived using the relatively
bright part of the galaxy LF, differential effects between the
field and cluster galaxy LFs should be small.
Using the three different LFs listed in YL99 produces less than
$\sim10$\% variations in the \bgc~values.
Furthermore,  adjusting $Q$ by $\pm$30\% does not change the
relative \bgc~values of the high and low redshift clusters
significantly.

The resultant \bgc~measurements are listed in Table 1, along with
their estimated Abell richness classes.
The \bgc~values, in units of $h_{50}^{-1.8}$ Mpc$^{1.8}$ 
can be roughly calibrated to the Abell richness
class scale as follows (YL99): Abell 0: $400<B_{gc}<800$;
Abell 1: $800<B_{gc}<1200$; and Abell 2: $1200<B_{gc}<1600$.

\begin{figure}[ht] \figurenum{1}\plotone{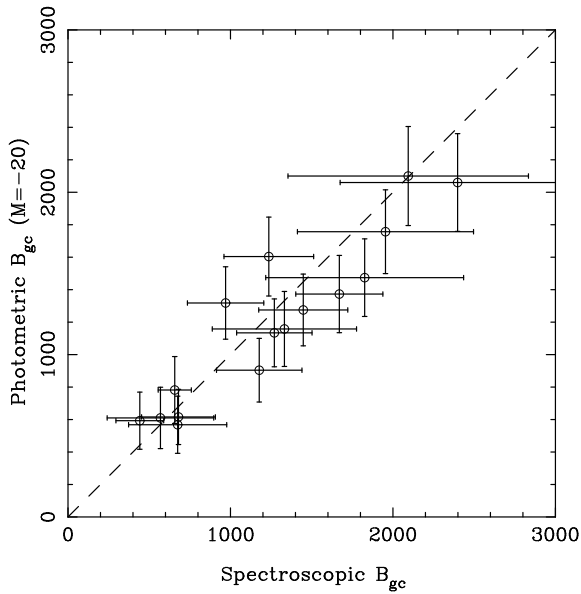} 
\caption{
Photometric \bgc~vs spectroscopic \bgc, showing that richness can
be estimated robustly using photometric data alone.
The dashed line has a slope of 1.
}\end{figure}

\subsection{Spectroscopic \bgc}

The CNOC1 sample offers the unique opportunity to 
verify the robustness of photometrically determined richness
parameters with those determined using a redshift sample.
This allows us to test the validity and accuracy of 
statistical background corrections.
We determine the \bgc~parameters for the clusters using
the same basic procedure as that for the photometric measurements.
Here, galaxies considered to be in the cluster, based on the
redshift range established for each cluster in Carlberg et al.~(1996),
are summed according to their magnitude and geometric weights (Yee et al.~1996)
to arrive at the estimated net counts.
In general the spectroscopic \bgc's are determined using 
absolute magnitude limits brighter than --20 (due to the relatively bright 
spectroscopic limits), and hence have larger uncertainties
despite not having to correct for background counts.

The \bgc~values derived from the spectroscopic sample
are plotted versus the photometric \bgc~in Figure 1.
There is a very strong correlation between the two measurements
with a slope close to the expected unity.
The scatter in the relation is consistent with the estimated 
uncertainties.  From this we can conclude 
that \bgc~values determined
using photometric data with statistical background count corrections are 
reliable estimates of cluster richness, even to redshifts of $\sim0.6$.
This further demonstrates that photometric \bgc~is a robust
richness parameter.

\section{Correlations of Richness and Cluster Global Properties } 

\subsection {Expected Scaling Relations}
The richness parameter \bgc~scales with
the net counts of galaxies to some fixed luminosity within a certain radius.
By making the simplifying assumption that 
galaxy clusters have similar galaxy luminosity functions at the
bright end (e.g., L\'opez-Cruz 1997 found that $M^*$ for the
LOCOS sample has a scatter of $<0.25$ mag), galaxy spatial distributions, and
mass-to-light ratios (e.g., Carlberg et al.~1996), \bgc~is then
expected to track the mass of clusters,
and essentially measures the scaling of the spatial 
distribution of galaxies around the center of the cluster.
Hence, \bgc~scales as the mass of clusters at a {\it fixed} proper radius.
However, this is not an interesting correlation, as mass measures
such as $M_{vir}$, the virial mass, or $M_{200}$, (the mass interior to
 $r_{200}$, where the average mass density is 200$\rho_c$), are 
connected more directly to observational quantities such as 
$\sigma_1$ and \Tx.
We can derive the links between \bgc~and these properties using
the following scaling relations.

Assuming a correlation function form of $\xi(r)=B_{gc}r^{-\gamma}$ and
that galaxies trace the total mass, we expect the mass density
$\rho\sim r^{-\gamma}$,
and hence the mass interior to some radius $r$, 
$M(r)\propto B_{gc}r^{-\gamma+3}$.
The mass of a cluster at some fixed fiducial radius $r'$ is related
to the virial mass $M_{vir}$ at radius $r_{vir}$ by
$M_{r'}\propto M_{vir}r_{vir}^{\gamma-3}$.
In the virialized region,  $r_{vir}\propto\sigma_1$
(e.g., see Carlberg et al.~1996)  and
$M_{vir}\propto \sigma_1^3$.
Hence, we expect the simple relationship:
$$B_{gc}\propto\sigma_1^{\gamma},\eqno(1)$$
and
$$B_{gc}\propto M_{vir}^{\gamma/3}. \eqno(2)$$
Note that the same relationship holds for $M_{200}$.
Since for a virialized system $M_{vir}\propto T_{\rm X}^{3/2}$
(e.g., see Horner et al.~1999), we also expect:
$$B_{gc}\propto T_{\rm X}^{\gamma/2}. \eqno(3)$$
Finally, another important parameter for clusters is $r_{200}$, which
provides one with a physical size scaling of clusters of various
richness.  Since $r_{200}\propto\sigma_1$, 
we have:
$$B_{gc}\propto r_{200}^{\gamma}. \eqno(4)$$

We note that for computing  \bgc~we have chosen to use 
$\gamma=1.8$, primarily to allow for a direct comparison with 
the galaxy-galaxy correlation amplitude.
Analyses of correlations between galaxies and clusters have shown
that $\gamma$ is typically steeper than that for galaxy-galaxy
correlation functions, varying from 1.8 to 2.4 (e.g., 
Lilje \& Efstathiou 1988; Moore et al.~1994; Croft, Dalton, \&
Efstathiou~1999), 
with the low value corresponding to the correlation 
between clusters and IRAS galaxies.
However, for the purpose of using the correlation amplitude
as a measure of richness, a shallower $\gamma$ of 1.8 is likely
the more appropriate one, due to the small cluster-centric radius
used for the counting of galaxies.
The $\gamma$'s derived for galaxy-cluster correlations are generally
measured over scales of 10 $h_{50}^{-1}$ Mpc or more, compared to
 0.5 $h_{50}^{-1}$ Mpc used for the estimate of \bgc.
These correlation function slopes essentially reflect the
dependence of cluster profiles at large radii.
Profiles such as NFW (Navarro,
Frenk, \& White 1997) and Hernquist models (Hernquist 1990)
typically have a form of $r^{<-2}$ at large $r$, 
and asymptotically approach $r^{<-3}$, similar to the measured
correlation functions.
At smaller radii, however, cluster profiles in general have flattening
slopes, possibly leading to a core.
At the typical 0.5 $h_{50}^{-1}$ Mpc where \bgc~is measured
(equivalent to $\sim 0.25r_{200}$),
a $\gamma$ of 1.8 is a reasonable representation.

\begin{figure}[ht] \figurenum{2}\plotone{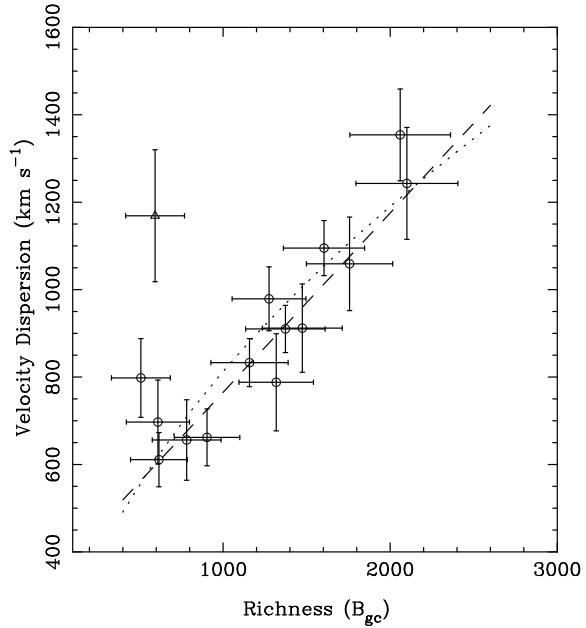} \caption{
Line-of-sight velocity dispersion vs richness as parametrized by \bgc.
The outlying point on the left hand edge (open triangle) is MS1445+22.
The dashed line is the best linear fit  with
MS1445+22 removed, while the dotted line is the power-law fit.
}\end{figure}

\begin{figure}[ht] \figurenum{3}\plotone{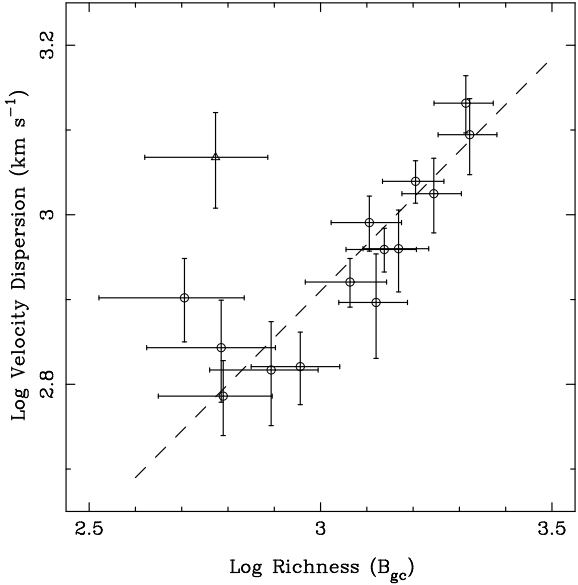} 
\caption{
Velocity dispersion vs \bgc~in logarithmic scale.
The dashed line is the best fitting function using BCES.
MS1455+22, omitted from the fit, is indicated by an open triangle.
}\end{figure}

\subsection{Correlations of Richness with Global Optical Properties}
In this section we examine the correlation of
the richness of the CNOC1 clusters with quantities derived
from optical data, such as velocity dispersion, mass, and $r_{200}$.
The main motivation in deriving these dependences is to use \bgc,
an easily measurable parameter using just imaging data,
as a predictor of the other quantities.
Hence, we fit expressions of these quantities as a function of \bgc.

The \bgc~parameter provides an excellent correlation with velocity dispersion.
We plot in Figure 2 the correlation of velocity dispersion (from Carlberg,
Yee, \& Ellingson 1997) 
 vs the photometric \bgc~richness parameter using linear scaling, 
and in Figure 3 using logarithmic scaling.
In the plots, there is one cluster which stands out as an outlier:
MS1455+22, which has too large a velocity dispersion for its
richness.
We exclude MS1455+22 from all our fits, but include it in the plots.
We further discuss the case of MS1455+22 in \S5.
Using the bisector BCES estimator fitting routine (which allows for
measurement errors in both variables;
Akritas \& Bershady 1996) in logarithmic space,
we obtain a fit plotted as a dashed line in Figure 3:
  $${\rm log}\,\sigma_1= (0.55\pm0.09)\,{\rm log} B_{gc} + (1.26\mp0.30),
\eqno(5)  $$ 
where $\sigma_1$ is in units of km s$^{-1}$, and \bgc, in units of
$h_{50}^{-1.8}$Mpc$^{1.8}$.
The fit is equivalent to $B_{gc}\propto\sigma_1^{1.8\mp0.2}$,
entirely consistent with the expected relation of eqn 1 for $\gamma=1.8$.
We note that varying $\gamma$ in the derivation of \bgc~produces
values that are simply scaled by the integration constant $I_\gamma$
(defined in Longair \& Seldner 1979), and does not
affect the power-law slope of the fit.
Hence, the correlation between velocity dispersion and \bgc~indicates
that the $\gamma$ used is consistent with the galaxy density
slope near 0.5$h_{50}^{-1}$ Mpc.

Since the data span a factor of only  $\sim3$ in richness, we also perform
a linear fit over the data range to provide a convenient formula for
estimating the velocity dispersion from \bgc:
 $$\sigma_1 =(416\pm61) {B_{gc}\over 1000}+ (340\mp90) {\rm~km~s}^{-1}, \eqno(6)$$ 
in the range of $500<B_{gc}<2500$.
In Figure 2, we plot both the linear and power-law fits between
$\sigma_1$ and \bgc.

In Figures 4 and 5, we plot in logarithmic scale the
relationship between \bgc~and $M_{200}$ and $r_{200}$,
with the following best BCES fits:
$${\rm log }\, M_{200}=(1.64\pm0.28)\,{\rm log} B_{gc} + (10.05\mp0.89),
\eqno(7)$$
and
$${\rm log }\,r_{200}=(0.47\pm0.16)\,{\rm log} B_{gc} - (1.05\mp0.48), \eqno(8)$$
where $M_{200}$ is in units of $h_{50}^{-1} 
M_\odot$, and $r_{200}$, in units of $h_{50}^{-1}$ Mpc.
These correlations are equivalent to $B_{gc}\propto M_{200}^{0.61\pm0.10}$, and
$B_{gc}\propto r_{200}^{2.1\pm0.7}$, in excellent agreement with the
expected values of 0.60 and 1.8 for $\gamma=1.8$.
We note that the scatter in the $r_{200}$ versus \bgc~correlation is
rather large, and significantly greater than the error bars.

\begin{figure}[ht] \figurenum{4}\plotone{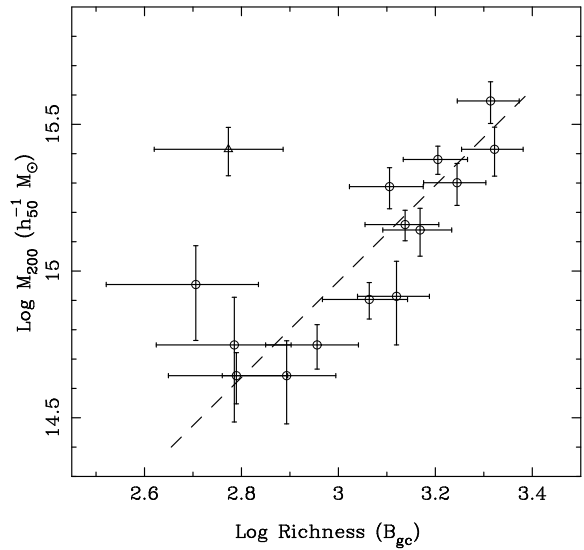} 
\caption{
$M_{200}$  vs \bgc. 
The dashed line shows the best fitting power law.
MS1455+22 is indicated as an open triangle.
}\end{figure}

\begin{figure}[ht] \figurenum{5}\plotone{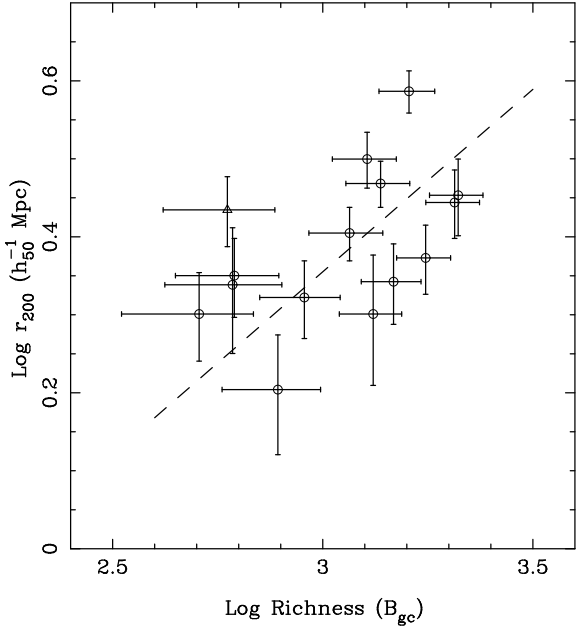} \caption{
$r_{200}$ vs \bgc. The dashed line shows the best fitting power law.
MS1455+22 is indicated as an open triangle.
}\end{figure}

\subsection{Correlations of Richness with X-ray Properties}
X-ray temperature and luminosity have been used as key measures
of cluster properties.
From simple physical models, it is expected that these quantities
scale with properties such as mass and velocity dispersion.
Given the excellent correlations of the richness parameter \bgc~with
mass and velocity dispersion, the \bgc~parameter should 
also be a useful predictor for X-ray properties.

Figure 6 presents the correlation between \Tx~and \bgc.
The bisector fit for the relation is:
$${\rm log}\,T_{\rm X} = (0.78\pm0.13)\,{\rm log}B_{gc}-(1.59\mp0.40),\eqno(9)$$
where \tx~is in units of keV.
Several of the EMSS clusters have their total fluxes affected by
cooling flows in the cluster core (Lewis et al.~1999), which may
affect the temperature measurement. 
The four clusters expected to be most affected by cooling flows
(MS0839+29, MS1224+20, MS1358+62, and MS1445+22) are
 marked by separate symbols on Figure 6.
We repeat the fit with these objects removed (leaving 10 clusters),
producing essentially an identical result.
The \tx--\bgc~scaling of Eqn (9) corresponds 
to \bgc$\propto T_{\rm X}^{1.28\pm0.21}$,  almost 2 
sigma higher than the expected power-law exponent of 0.90 for $\gamma=1.8$.
The expected relationship, however, is predicated 
on $\sigma_1\propto T_{\rm X}^{0.5}$.
Different investigators have obtained different dependences between $\sigma_1$ 
and \tx, with  exponent
values  of $0.62\pm0.04$ (Girardi et al.~1998), 
$0.67\pm0.09$ (Wu et al.~1999), and $0.53\pm0.04$ (Horner et al.~1999).
Figure 7 shows the  $\sigma_1$ vs \Tx~relation for the CNOC1 sample.
The best fit gives a relationship of $\sigma_1\propto T_{\rm X}^{0.66\pm0.13}$,
closer to the steeper of the current literature results.
Assuming such a \Tx-$\sigma_1$ relationship, we would 
expect a steeper correlation, \bgc$\propto T_{\rm X}^{1.20}$, which
is well within one sigma of the result obtained.

\begin{figure}[ht] \figurenum{6}\plotone{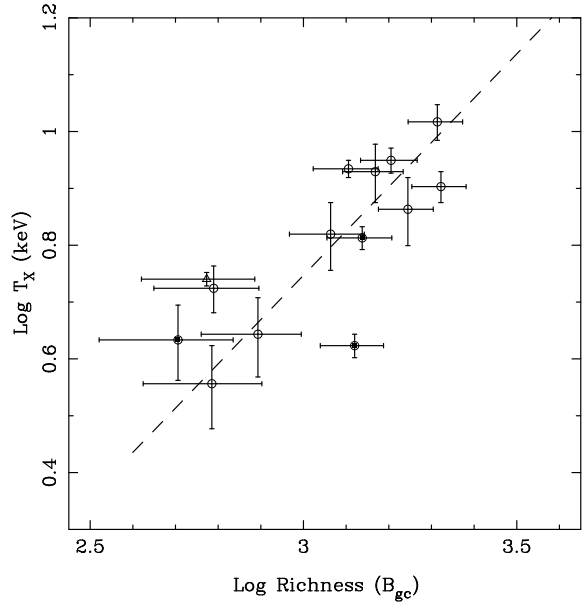} \caption{
X-ray temperature vs \bgc. The dashed line shows the best 
fitting power law using all data except MS1455+22. 
 MS1455+22 is indicated as a triangle.
Three other clusters with prominent cooling flows are indicated by
imbedded solid circles.
These three clusters, along with MS1455+22, produce a
significant fraction of the scatter in the relation.
Note that the most discrepant point is not MS1455+22, but
MS0839+29, which is the point well below the fitted relation.
}\end{figure}

\begin{figure}[ht] \figurenum{7}\plotone{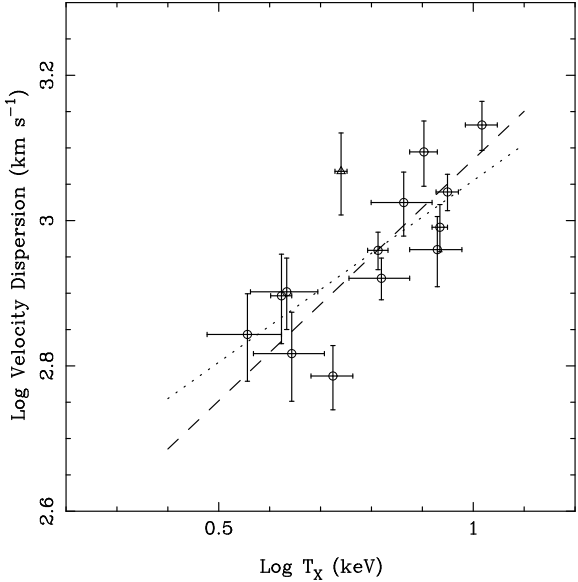} \caption{
Velocity dispersion vs \Tx. The dashed line shows the best fitting power law.
The dotted line has a slope of 0.5, the expected scaling between
\Tx~and $\sigma$.
}\end{figure}

The relationship between the bolometric \lx~and \bgc~is plotted in Figure 8.
The best BCES fit is:
$${\rm log}\,L_{\rm X} = (1.84\pm0.24)\,{\rm log}B_{gc}-(4.48\mp0.75),\eqno(10)$$
where \lx~is in units of 10$^{44}$ $h_{50}^{-2}$ erg s$^{-1}$.
The correlation is equivalent to $B_{gc}\propto L_{\rm X}^{0.54\pm0.07}$.
We note that careful attention must be given to the various
corrections required for the determination
of the bolometric \lx, which may introduce a change of a factor 
two or more to the result (e.g., see Ellis \& Jones 2002).
A considerably larger scatter in the correlation is obtained using
existing values from the literature (e.g., from the compilation of
Wu et al.~1999; or from the EMSS measures from Gioia \& Luppino 1994).

\begin{figure}[ht] \figurenum{8}\plotone{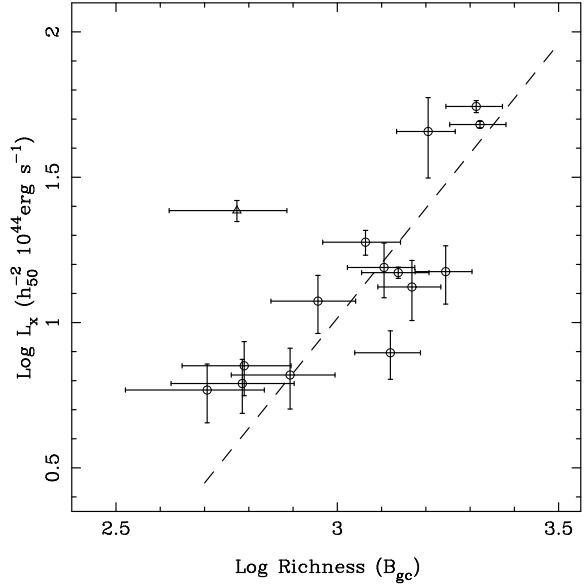} \caption{
\Lx~vs \bgc. The dashed line shows the best fitting power law.
MS1455+22 is indicated by an open triangle.
}\end{figure}

\section{Discussion}
\subsection{\bgc~as an Estimator of Cluster Properties}
We have found that for the CNOC1 clusters strong
correlations exist between the optical richness and other directly
measured quantities such as velocity dispersion and \tx,
especially after a single discrepant cluster, MS1455+22, is
removed.
Here, we examine the goodness of fit tests for the various relationship
to obtain an indication of how well \bgc~can be used as a predictor
of other global properties of clusters.

The  rms scatter of the CNOC1 sample velocity dispersions from the
best fitting power-law relation, excluding MS1455+22, is
105 km s$^{-1}$, or $\sim$12\% of the mean value. 
The reduced $\chi^2$ of the fit is 0.65. 
We note that, because there are uncertainties in the data in both
the $x$ and $y$ axes, $\chi^2$ is estimated using the effective
variance: $\sigma_{eff}^2=\sigma_y^2+a^2\sigma_x^2$, where $a$
is the slope of the power-law fit.
By including MS1455+22, the rms deviation is increased
to 176 km s$^{-1}$ (or 20\%), with the reduced $\chi^2$ increased to 1.3.
Similar results are obtained when using the linear relation
of Eqn 6, but with larger reduced $\chi^2$ of 1.4 and 2.4 for
with and without MS1455+22, respectively.
These somewhat higher $\chi^2$ values are probably a more robust
indication of the goodness of fit, as the effective variance used
is strictly correct only for a linear fit.
The relatively small $\chi^2$ and fractional scatter suggests that, 
with the exception of
MS1455+22, the $\sigma_1$ vs richness relation is consistent with
little or no intrinsic scatter beyond the measurement 
uncertainties, and \bgc~can be used as a reasonably
accurate estimator of velocity dispersion.

For the dynamically  derived quantities, such as mass and $r_{200}$,
the scatter of the data points around the best fitting relation
is significantly larger than that of $\sigma_1$.
The rms scatter in $M_{200}$ from the best fitting power law
is 4.4$\times 10^{14} h_{50}^{-1} {\rm M}_\odot$
 amounting to 31\% of the average value of the sample; while
for $r_{200}$, the scatter is 0.52 $h_{50}^{-1}$ Mpc, or 21\% of 
the average value.
The reduced $\chi^2$'s are 0.83 and 2.0 for the $M_{200}$ and $r_{200}$
relation, respectively.

In general the correlations between richness and X-ray quantities
have somewhat larger $\chi^2$'s.
For the \tx~vs \bgc~relationship, we obtain for the 13 data points
(excluding MS1455+22) a
scatter of 1.36 keV relative to the best-fit power law, about 21\% of
the mean \tx~of the sample.  However,
a relatively large reduced $\chi^2$ of $\sim2.1$ is obtained for the best fit,
indicating that there may be a significant intrinsic variance in 
the \tx~vs richness relationship.  
Or alternatively, this scatter could be an indication that
there may be systematic uncertainties in the \tx~determinations
not taken into account by the error bars listed in the literature.
The correction of \tx~for cooling flows and departures from isothermal
conditions is a complicating issue
(e.g., see Allen 1998; Lewis et al. 1999),
and may contribute to the larger $\chi^2$.
Removing the three additional clusters with the most prominent cooling flows,
a scatter of 1.2 keV about the best fitting relation with a 
reduced $\chi^2$ of 1.2 is obtained, indicating that cooling flows in clusters
may be an important source of the larger scatter seen in this relationship.

The correlation between \lx~and \bgc~
has the largest scatter.
The reduced $\chi^2$ (excluding MS1455+22)
of the best fit  is $\sim1.5$, and the rms
scatter of the data from the fit is 47\%.
Removing the three additional cooling flow prominent clusters
improves the goodness of fit only marginally.

\subsection {\bgc, \tx, and \lx~as Estimators of Dynamical Mass}

Both \tx~and \lx~have often been used as estimators or proxies
 for the mass of clusters
in the study of mass functions of clusters (e.g., Oukir \& Blanchard
1997; Viana \& Liddle 1999; Henry 2000).
Hence, it is useful to examine the correlations between \tx~and \lx~with
the dynamically derived $M_{200}$ of this sample, and compare their
relative merits as estimators of mass to that of \bgc. 

Figure 9 shows the excellent correlation between \tx~and $M_{200}$, with
the best power-law fit:
$${\rm log}\,M_{200} = (2.01\pm0.33)\,{\rm log}T_{\rm X}
+(13.47\mp0.29),\eqno(11)$$
where $M_{200}$ is in units of $h_{50}^{-1}M_\odot$, and \tx~in units of keV.
We note that MS1455+22 is a significant outlier also in this
correlation -- it has too high a virial mass for its \tx.
For simple virialized systems, we expect 
$M_{200}\propto T_{\rm X}^{1.5}$.
The relation derived here has a higher exponent, but is marginally
consistent at 1.5 sigma with the scaling expected for virialized systems.
The result, however, is consistent with
a number of studies of X-ray clusters showing
that $M\propto T_{\rm X}^{\sim1.7}$ (e.g., Ettori \& Fabian 1999;
Nevalainen, Markevitch, \& Forman 2000).
We note that these studies use mass determined using X-ray
data, whereas the CNOC1 relation is for dynamically determined
mass.
The similar results indicate the consistency in the two mass determination
techniques.

We find \lx~also correlates well with $M_{200}$, as shown in Figure 10, 
with:
$${\rm log}\,M_{200} = (0.89\pm0.11)\,{\rm log}L_{\rm X}
+(14.03\mp0.14),\eqno(12)$$
where \lx~is in units of $10^{44} h_{50}^{-2}$ erg s$^{-1}$.
The correlation is consistent with a linear relationship between
\lx~and mass.
The self-similar isothermal model with pure bremsstrahlung radiation
and viral equilibrium leads to the scaling relation of $T_{\rm X}\propto
L_{\rm X}^{0.5}$, (e.g., see Arnaud \& Evrard 1999).
Assuming the relation from simple virialized systems,
$M_{200}\propto T_{\rm X}^{1.5}$, we would expect
$M_{200}\propto L_{\rm X}^{0.75}$, which is within 1.5 sigma of the
relation obtained here with the CNOC1 sample.
If we use the measured $M_{200}\propto T_{\rm X}^{2.01\pm0.33}$,
we would predict $M_{200}\propto L_{\rm X}^{1.0\pm0.17}$, in good agreement
with the measured correlation.
In investigations of available X-ray data of galaxy clusters,
it has been found in general that $T_{\rm X}\propto L_{\rm X}^{<0.5}$
(e.g., Markevitch 1998; Allen \& Fabian 1998; Arnaud \& Evrard 1999; 
see Babul et al.~2002 for detailed modeling of the \tx--\lx~relation).
The current data set produces a scaling of $T_{\rm X}\propto 
L_{\rm X}^{0.42\pm0.06}$, consistent with the correlations found
for high temperature systems (e.g., Allen, Schmidt, \& Fabian 2001;
Babul et al.~2002).

The CNOC1 sample has shown that all three quantities: richness,
x-ray temperature, and x-ray luminosity, correlate well with the
derived dynamical mass.
Here, we compare the relative merits of these quantities as predictors 
of the mass of galaxy clusters.
The \bgc~parameter produces a scatter in the predicted
$M_{200}$ around 31\%, with a reduced $\chi^2$ of 0.83.
This compares with a scatter in $M_{200}$ of 29\% and a reduced 
$\chi^2$ of 1.6 for the \tx--$M_{200}$ relation, and a scatter 
of 35\% and a reduced $\chi^2$ of 2.8
for \lx, all computed without MS1455+22.
Thus, all three quantities can serve as good predictor of
the cluster dynamical mass to the level of $\sim30$\%, 
with \tx~and \bgc~being somewhat superior to \lx.
However, we note that the proper determination of the bolometric
\lx~is not a simple
procedure, and in fact {\it requires} the knowledge of \tx~and
the X-ray emission spatial profile.
X-ray luminosities derived without careful attention to spatial
and spectral distributions produce a much
poorer correlation with mass and other quantities than shown here.
Hence, \lx~is not as useful an
estimator of cluster mass as \tx~or \bgc.

\begin{figure}[ht] \figurenum{9}\plotone{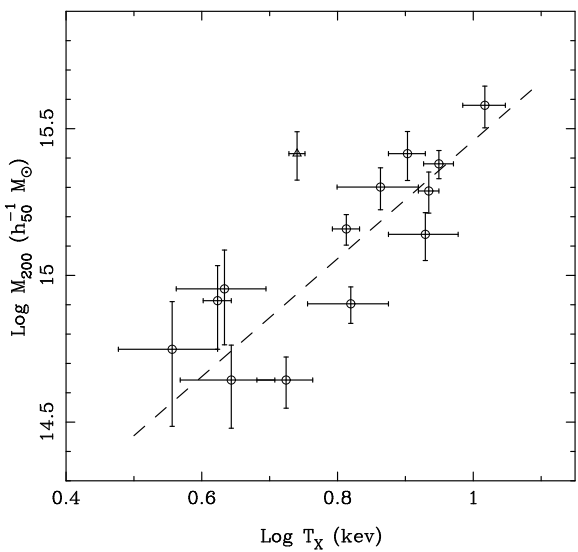} \caption{
Dynamically determined mass $M_{200}$~vs \tx. The dashed line shows the 
MS1455+22 is indicated by an open triangle.
}\end{figure}

\begin{figure}[ht] \figurenum{10}\plotone{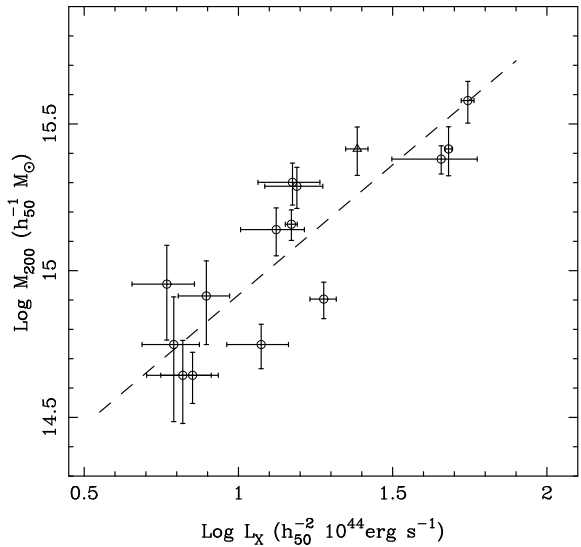} \caption{
Dynamically determined mass $M_{200}$~vs \lx. The dashed line shows the 
best fitting power law.
MS1455+22 is indicated by an open triangle.
}\end{figure}

That optical richness can predict the mass of clusters as accurately as
X-ray properties is not entirely surprising.
Carlberg et al.~(1996) showed that for the CNOC1 sample
$L_{200}$, the luminosity 
within $r_{200}$, is well correlated with velocity dispersion.
Girardi et al.~(2002), using a large sample of low-redshift clusters, 
also found a good correlation between the $B$ band luminosity within the
viral radius and virial mass with $M_v\propto L_B^{\sim1.2}$.
However, neither of these luminosity measurements
 is a simple quantity to derive directly
from photometric data alone, as it requires the knowledge of the
virial radius or $r_{200}$.
The derivation of \bgc~does not require the knowledge of the size
of the cluster because it assumes a fixed form of the density function.
Furthermore, such an assumption allows one to estimate the 
characteristic size of the cluster (e.g., $r_{200}$).
The correlation between \bgc~and dynamical mass derived here spans only
a small mass range of about a factor of 10.
There is a significant amount of  both observational and
theoretical evidence that, at least at lower masses, 
the assumption of a constant mass-to-light ratio is not a valid one
(e.g., Carlberg et al.~2001; Girardi et al.~2002; Bahcall et al.~2000).
This will affect the \bgc--mass correlation, and
thus, the scaling relations between \bgc~and $\sigma_1$, $M_{200}$,
and $r_{200}$ derived in \S4 probably cannot be extrapolated into 
a mass regime much smaller than $10^{14} M_\odot$.
Further calibrations of these relations are needed using a larger
mass range.

\subsection {The Case of MS1455+22}
The galaxy cluster MS1455+22 is a significant outlier to most of
the correlations discussed in \S4:
It has either too low a \bgc~value (by about a factor of three)
for its velocity dispersion, 
too high a velocity dispersion (by about a factor of two) for its richness,
or a combination of both.
In the distribution of mass-to-light ratios of the CNOC1 clusters,
MS1455+22 also stands out with an anomalously large (at the 2$\sigma$
level) value of 807 $h_{100}$ M$_\odot$/L$_\odot$,
compared to the mean of 295 $h_{100}$ M$_\odot$/L$_\odot$ for the whole
sample (Carlberg et al.~1996).
In many ways it is not surprising to find outliers to the correlations.
Galaxy clusters are complex systems; differential evolution and events such as
major mergers should produce intrinsic variances to these relations.
Hence, it is of interest to examine clusters that are significantly
discrepant from the average population.

MS1455+22 itself appears to be a very evolved cluster with a 
luminous and large central galaxy,  a well-formed red sequence with
a relatively small blue fraction (0.15; Ellingson et al.~2000),
and regular X-ray morphology (Lewis et al.~1999).
In addition, it has one of the largest cooling flow measurements
in the X-ray (Lewis et al.~1999).
The fact that the spectroscopic and photometric \bgc's agree within
the uncertainties indicates that the \bgc~value is not affected by
an unusually low background count.
If \bgc~is anomalously low compared to other clusters, 
one possible explanation is that MS1455+22 is in a more advanced
state of evolution than the others, as suggested by
the large cooling flows, the strong red sequence, and the dominant
central galaxy.
The dominant central galaxy of MS1455+22 could contribute to
the low \bgc~value by having cannibalized a larger number of
its bright members, leaving a lower net cluster galaxy counts.
The more evolved galaxy population could produce a lower average
luminosity for the galaxies, which again may produce a smaller \bgc~value.
However, the small number statistics of the redshift sample
are not sufficient for us to
draw definitive conclusions regarding the LF of the galaxies.

The velocity dispersion for MS1455+22 was determined using 51 cluster
galaxies (Carlberg et al.~1996), and hence is expected to be relatively
robust.
Nevertheless, there are indications that the 1196 km s$^{-1}$ value
could be somewhat high. 
One indication can be found in the  $\sigma_1$ vs \tx~plot where
MS1455+22 has a high velocity dispersion for its \tx, while
in the \tx~versus \bgc~correlation MS1455+22 is not particularly
discrepant.
However, a typical cooling flow correction may bring \tx~to a
better agreement with the velocity dispersion.
Second, Borgani et al.~(2000) applied several different methods to the CNOC1
data in determining the velocity dispersion, showing that the
results, as a sample, are consistent.
However, the different methods in Borgani et al.~produced consistently
lower $\sigma_1$ for MS1455+22, ranging from 962 km s$^{-1}$ to
1033 km s$^{-1}$.
(We note that using the different velocity dispersion derivations
 in Borgani et al.~produce essentially the same correlations with \bgc,
but with somewhat larger scatter, with MS1455+22 being a less
discrepant point.)

Finally, in comparing the masses determined using  X-ray data and weak lensing,
Lewis et al.~(1999) found that while the dynamical mass of MS1455+22 is
consistent with the weak-lensing mass, it
is 1.34 times that of the X-ray mass.
Since weak lensing tends to over estimate cluster mass on the average
(Lewis et al.~1999), 
this may suggest that the dynamical mass of MS1455+22 is somewhat high.

In summary, current data do not provide a single clear 
explanation of the discrepancy of MS1455+22 in an otherwise tight 
correlation of $\sigma_1$ with \bgc. 
There is some evidence that the velocity dispersion may be somewhat high,
but not sufficient to account for a factor of two.
An intriguing suggestion is that the \bgc~value is significantly low
due to the cluster being in an evolutionary state different from the
others, with more evolved galaxies and a very dominant central galaxy.

\subsection {Comparisons with Abell Clusters}
The CNOC1 sample is one of the most 
homogeneous samples for investigating the correlations of 
various properties.
The parent sample of the CNOC1 sample is the EMSS (with one
additional Abell cluster with similar X-ray properties), which at the
time when CNOC1 was carried out was the largest and most
homogeneous X-ray cluster catalog.
Both the optical and X-ray data (Yee et al.~1996, and Lewis et al.~1999)
were obtained in a systematic manner and predominantly from a single source.
Nevertheless, the CNOC1 sample is very small, and it would
be instructive to see if a larger sample of clusters with
data coming from more heterogeneous sources follow these
correlations.
Such a comparison will give confidence on the use of these
relationship as predictors of various cluster properties.

An excellent set of clusters for such a comparison is the LOCOS 
sample of L\'opez-Cruz and Yee (see L\'opez-Cruz 1997 and YL99).
The LOCOS is an imaging survey of 46 low-redshift Abell clusters chosen
from the compilation of X-ray detected Abell clusters from
Jones \& Forman (1999).  The photometric catalogs were
created using a procedure identical to that of the CNOC1 survey,
with the exception that the $R_c$ filter band was used instead of Gunn $r$.
We remove the small number of Abell richness class 0 clusters from 
the LOCOS samples, as the survey sampled fairly only clusters with
richness class $\ge1$ (see YL99).
Data for the velocity dispersion, \Tx, and \Lx~are obtained
from the literature.

\begin{figure}[ht] \figurenum{11}\plotone{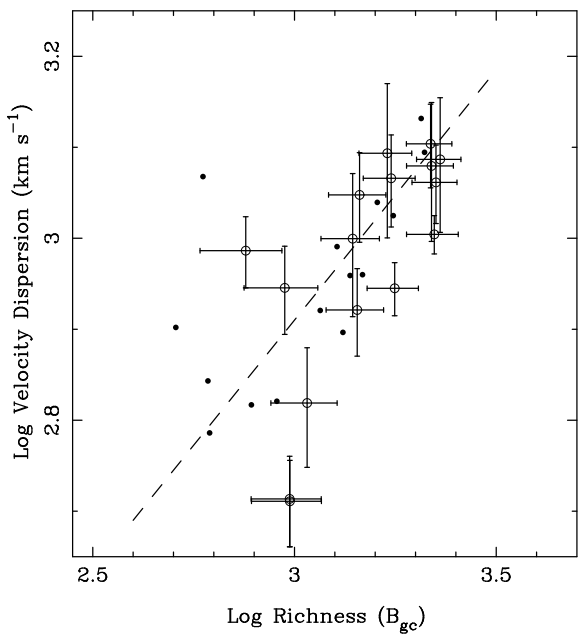} \caption{
Velocity dispersion versus \bgc~for Abell clusters (open circles).
Also plotted for comparison are the CNOC1 clusters (solid circles
without error bars)
and the power-law fit from the CNOC1 clusters.
}\end{figure}

For the correlation with velocity dispersion, we use data 
from Fadda et al.~(1996) and Girardi et al.~(1993). 
YL99 showed a $\sigma_1$ vs \bgc~plot using all clusters with
$\sigma_1$ determined using 10 or more velocities, producing
a correlation with a relatively large scatter.
Here we limit our $\sigma_1$ data to those comparable to the
CNOC1 survey, resulting in 16 clusters with $\sigma_1$ measured using
25 or more velocities. 
The result is shown in Figure 11.
Plotted for comparison are both the CNOC1 points and the
best CNOC1 fit.
It can be seen that the Abell data points fall on the relation
on average,
albeit with a somewhat larger scatter than the CNOC1 clusters,
with the most discrepant clusters being at lower richness.
The deviation of the Abell cluster $\sigma_1$'s from that
predicted by their \bgc~has a rms scatter of 170 km s$^{-1}$,
and an offset of --55 km s$^{-1}$.
The scatter is equivalent to 17\% of the average velocity dispersion
of the sample.
The scatter is larger than the CNOC1 sample with MS1455+22 removed,
but slightly smaller than the full CNOC1 sample.
Thus, using a completely different sample, selected with different
criteria and with velocity dispersions obtained from different
sources, we have found that the \bgc~values can predict the
velocity dispersion with an uncertainty of better than $\sim 20$\%
(about 150 to 200 km s$^{-1}$),
and similar to that indicated by the CNOC1 sample itself.
This demonstrates that the richness parameter \bgc~is a useful
predictor of the velocity dispersion of a cluster, and hence an
estimator of its mass.
The two low-velocity dispersion Abell clusters which show the largest
deviations from the fitted relation may be an indication of the
nature of the Abell cluster sample which favors finding clusters with
projection contaminations.

For the correlation between richness and X-ray properties,
we use the compilations of \tx~and \lx~from Wu et al.~(1999).
Figure 12 plots the \tx~vs \bgc~relationship, with the
CNOC1 data points and best fit indicated.
The Abell cluster points fall slightly below the CNOC1 relation,
with a similar slope and a larger scatter, especially
towards the low \tx~direction.
However, we note that the most discrepant points are
also the ones with the largest \tx~error bars, and most clusters are
within 2$\sigma$ of the regression line from the CNOC1 data.

\begin{figure}[ht] \figurenum{12}\plotone{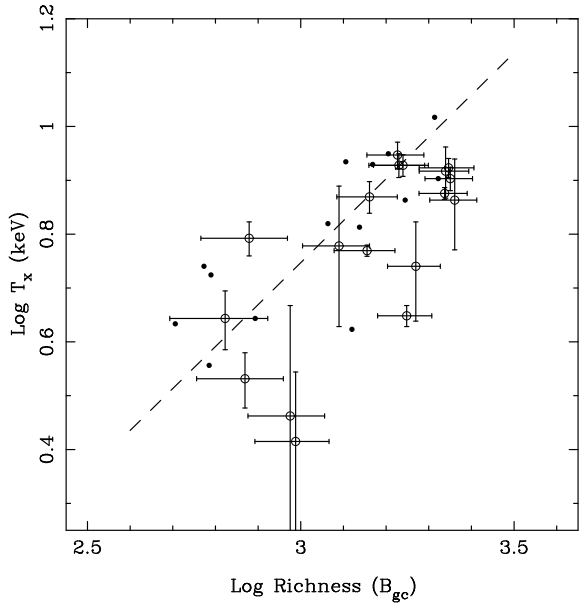} \caption{
\tx~versus \bgc~for Abell clusters (open circles).
Also plotted for comparison are the CNOC1 clusters (solid circles
without error bars)
and the power-law fit from the CNOC1 clusters.
}\end{figure}

\begin{figure}[ht] \figurenum{13}\plotone{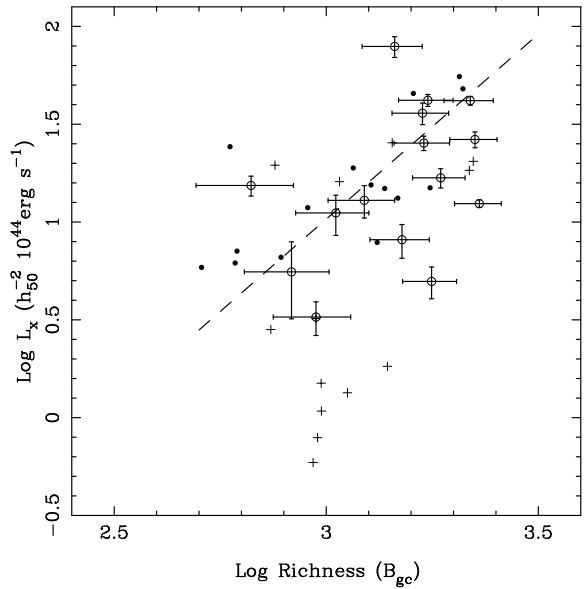} \caption{
X-ray luminosity versus \bgc~for Abell clusters from YL99 with
$z>0.065$ (open circles).
Also plotted for comparison are the CNOC1 clusters (solid circles
without error bars)
and the power-law fit from the CNOC1 clusters.
Abell clusters with $x<0.065$ are also plotted, without error bars,
as crosses.
}\end{figure}

We find that the \lx~vs richness~relationship for the
LOCOS sample has the largest scatter of all the
correlations discussed.
When all LOCOS clusters with available \lx~from Wu et al.~(1999)
are plotted, there is only a very rough suggestion of
a correlation (Figure 13).
A closer examination shows clearly that the magnitude
of the scatter in the data is a strong function of
the redshift of the cluster, with the
largest scatter belonging to clusters at the 
low end of the redshift range of the sample.
On Figure 13, we plot selectively the LOCOS data for clusters
with $z>0.065$.
The scatter is large, as indicated above, although
the points basically fall along the relation found
by the CNOC1 sample.
The $z<0.065$ clusters are plotted as crosses, and
it is clear that most of these objects have very
significantly lower \lx~than that predicted by the
CNOC1 relation.
The much poorer correlation likely indicates the inhomogeneous
nature of the \lx~data compiled by Wu et al. (1999).
There may also be systematic effects between the methods of
\lx~determination used for the CNOC1 sample and the Abell clusters.
Determining a consistent set of \lx~from a heterogeneous
database is complicated,
as data from different bands with different resolutions,
aperture sizes, and signal-to-noise ratios are used.
A number of $z<0.065$ clusters have \lx's which are a factor
of several lower than the expected value, which may arise from
incorrect aperture corrections due to the large
angular sizes of these very low-redshift clusters.
We note that well-studied clusters
at low redshift, such as Abell 1656 (Coma), in
fact do fall along the CNOC1 relation.

For both the \tx~and \lx~correlations with \bgc~there is a tendency 
for the Abell sample to fall below the CNOC1 regression line, indicating
a larger \bgc~relative to~\tx~or \lx.
This difference can arise from several sources, including simply the
systematics in the data sets used.
One interesting possibility is that this is an indication of
the relative biases between cluster samples selected using X-ray and optical
criteria.
The Abell sample is chosen via optical galaxy over-density, and hence
favors high \bgc~clusters, while the EMSS sample, being selected via
X-ray fluxes, preferentially finds high X-ray luminosity, hot clusters.
Large homogeneous samples at similar redshifts and selected using 
both methods are needed to investigate the existence of such biases.

\section{Summary} 

Using a small but homogeneous sample of clusters from the CNOC1
survey, we have demonstrated that the optical richness of the
clusters, measured by the galaxy-cluster center correlation
amplitude \bgc,  correlates well with various global properties of
the clusters.
In particular, the correlation of \bgc~with velocity dispersion is
excellent, with a rms scatter of only 100 km s$^{-1}$ when a single
outlier is removed, and 175 km s$^{-1}$ when the whole sample is
used.
The value of \bgc~as an estimator of velocity dispersion is verified
by using a sample of low-redshift Abell clusters, producing a similar
scatter around the best fit.

By extension, \bgc~is correlated with the dynamical mass (as represented
by $M_{200}$) and the virial size (as expressed by $r_{200}$) of
the clusters which are derived based on the velocity dispersion.
Similarly, \bgc~shows a strong correlation with \tx, and a somewhat 
more scattered correlation with \lx.
The forms of the correlations of the various properties with \bgc~are 
consistent with scaling laws
expected based on a simple spherical density distribution,
providing confidence in the existence of the underlying correlations.
We also compared the relative merits of \bgc, \tx, and \lx~as estimators
of the virial mass of the clusters.
We find that all three produce reasonably tight correlations, with
a scatter of $\sim30$\%.
However, the good correlation of \lx~with mass is predicated on proper
corrections to the \lx~determination, which require the knowledge of
\tx~and the spatial distribution of flux, 
making \lx~a less useful predictor of mass.
We note that the correlation of \tx~with dynamical mass is consistent 
with that found by other samples using masses determined via X-ray data.

The correlations investigated here demonstrate that
we should be able to use optical richness, a relatively inexpensive 
measurement, to estimate important attributes of a cluster.
This method is especially powerful in view of the current capability
of conducting large optical cluster surveys with improved
cluster finding techniques.
It allows samples of tens of thousands of clusters, for example, in
a 1000 square degree survey, to be characterized.
An important example of applications of these scaling relations
is the determination of the mass function of galaxy clusters
using large samples at different redshifts based on optical
photometric data, being carried out by the Red-Sequence
Cluster Survey out to $z\sim1$ (Gladders 2001;  Gladders \& Yee 2003).
A similar method of using total optical light to estimate cluster  mass
to derive the mass function of galaxy clusters at zero redshift has
been applied by Bahcall et al.~(2002) using imaging data from the SDSS survey.

The correlations discussed here are based on a small sample
of clusters over a relatively wide range of redshift of $0.17<z<0.55$.
To be able to use these scaling relations reliably for clusters
over a wide range of redshifts, it is important
to further calibrate these correlations covering a number of redshift
bins using comparable or larger samples from cluster surveys 
with well-understood selection functions.
These calibrations can be obtained
in the near future as cluster catalogs from large, deep cluster 
surveys are becoming available.

\acknowledgements
E.E. would like to acknowledge support from NSF grant AST-9617145 and
from NASA through the Chandra grant GO0-1079X.
The research of H.K.C.Y. is supported by an operating grant from
the Natural Sciences and Engineering Research Council of Canada.


\bigskip
\bigskip


\begin{references}
\reference{} Abraham, R.G., Yee, H.K.C., Ellingson, E., Gravel, P.,
Carlberg, R.G., \& Pritchet, C.J.  1998, ApJS, 116, 231
\reference{} Akritas, M.G. \& Bershady, M.A. 1996, ApJ, 470, 706
\reference{} Allen,  S.W. 1998, MNRAS, 296, 392
\reference{} Allen, S.W.  \& Fabian, A.C. 1998, MNRAS, 297, 63
\reference{} Allen, S.W., Schmidt, R.W., \& Fabian, A.C. 2001, MNRAS,
328, L37
\reference{} Anderson, V., \& Owen, F. 1994, AJ, 108, 361
\reference{} Arnaud, M. \& Evrard, A.E. 1999, MNRAS, 305, 631
\reference{} Babul, A., Balogh, M.L., Lewis, G.F., Poole, G.B. 2002,
MNRAS, in press, astro-ph/0109329
\reference{} Bahcall, N.A., Cen, R., Dav\'e, R., Ostriker, J.P.,
\& Yu, Q. 2000, ApJ, 541, 1
\reference{} Bahcall, N.A. et al.~2002, astro-ph/0205490
\reference{} Carlberg, R.G., Yee, H.K.C., Morris, S.L., Lin. H., Hall, P.B.,
Patton, D.R., Sawicki, M., \& Shepherd, C.W. 2001, ApJ, 552, 427
\reference{} Carlberg, R.G., Morris, S.L., Yee, H.K.C., 
\& Ellingson, E. 1997, ApJ, 479, L19
\reference{} Carlberg, R.G., Yee, H.K.C., \& Ellingson, E. 1997, ApJ, 478, 462
\reference{} Carlberg, R.G., Yee, H.K.C., Ellingson, E., Abraham, R.G.,
Gravel, P., Morris, S.L., \& Pritchet, C.J. 1996, ApJ, 462, 32
\reference{} Clowe, D., Luppino, G.A., Kaiser, N., Gioia, I.M. 2000,
ApJ, 539, 540
\reference{} Croft, R.A.C., Dalton, G.B., \& Efstathiou, G. 
1999, MNRAS, 305, 547
\reference{} Dickey, J.M. \& Lockman, F.J. 1990, ARA\&A, 28, 215
\reference{} Ellingson, E., Yee, H.K.C., Abraham, R.G., Morris, S.L., \&
Carlberg, R.G. 1998, ApJS, 116, 247, 262
\reference{} Ellis, S.C. \& Jones, L.R. 2002, MNRAS, 330, 631
\reference{} Ettori, S. \& Fabian, A.C. 1999, MNRAS, 305, 834
\reference{} Fadda, D., Girardi, M., Giuricin, G., Mardirossian, F., \&
Mezzetti, M. 1996, ApJ, 473, 670
\reference{} Gioia, I.M., Maccacaro, T., Schild, R. E., Wolter, A., 
Stocke, J.T., Morris, S.L., \& Henry, J.P. 1990, ApJS, 72, 567
\reference{} Gioia, I.M. \& Luppino, G.A. 1994, ApJS, 94, 583
\reference{} Girardi, M., Manzato, P. Mezzetti, M., Giuricin, G.,  \&
Limboz, F. 2002, ApJ, 569, 720
\reference{} Girardi, M., Borgani, S., Giuricin, G., Mardirossian, F., \&
Mezzetti, M. 2000, ApJ, 530, 62 
\reference{} Girardi, M., , Giuricin, G., Mardirossian, F., \&
Mezzetti, M. 1998, ApJ, 530, 62 
\reference{} Girardi, M., Biviano, A.,  Giuricin, G., Mardirossian, F., \&
Mezzetti, M. 1993, ApJ, 404, 38 
\reference{} Gladders, M.D. 2001, Ph.D.~Thesis, University of Toronto
\reference{} Gladders, M.D. \& Yee, H.K.C. 2001, AJ, 120, 2148
\reference{} Gladders, M.D. \& Yee, H.K.C. 2003, in preparation
\reference{} Grego, L., Carlstrom, J.E., Reese, E.D., Holder, G.P.,
Holzapfel, W.L., Joy, M.K., Mohr, J.J., \& Patel, S. 2001, ApJ, 552, 2
\reference{} Hernquist, L. 1990, ApJ, 356, 359
\reference{} Henry, J.P.  2000, ApJ, 534, 565
\reference{} Hoekstra, H., Franx, M., Kuijken, K., Squires, G. 1998,
ApJ, 504, 636
\reference{} Horner, D.J., Mushotzky, R.F., \& Scharf, C.A. 1999, ApJ, 520, 78
\reference{} Jones, C. \& Forman, W. 1999, ApJ, 511, 65
\reference{} Kent, S.M. \& Gunn, J.E. 1982, AJ, 87, 945
\reference{} Kim, R.  et al. 2002, AJ, 123, 20
\reference{} Levine, E.S., Schulz, A.E., \& White, M.  2002, ApJ, 577, 5
\reference{} Lewis, A., Ellingson, E., Morris, S., \& Carlberg, R.G. 1999,
ApJ, 517, 587
\reference{} Lilje, P.B. \& Efstathiou, G. 1988, MNRAS, 231, 635
\reference{} Lin, H., Yee, H.K.C., Carlberg, R.G., Morris, S., Sawicki, M.,
Patton, D., Wirth, G., \& Shepherd, C.W. 1999, ApJ, 518, 533 
\reference{} Longair, M.S., \& Seldner, M. 1979, MNRAS, 189, 433
\reference{} L\'opez-Cruz, O. 1997, Ph.~D.~thesis, University of Toronto 
\reference{} Markevitch, M. 1998, ApJ, 504, 27
\reference{} Moore, B., Frenk, C.S., Efstathiou, G., Saunders,W.
 1994, MNRAS, 269, 742
\reference{} Mushotzky, R.F., \& Scharf, C.A. 1997, ApJ, 482, L13
\reference{} Navarro, J., Frenk, C., \& White, S.M. 1997, ApJ, 490, 493
\reference{} Nevalainen, J, Markevitch, M., \& Forman, W. 2000, ApJ, 532, 694
\reference{} Oort, J.H. 1958, in La Structure et L'\'evolution de L'universe,
Onz\`eme Conseil de Physique, ed.R. Stoopes (Brussles : Solvay Inst.), 163
\reference{} Oukir, J. \& Blanchard, A. 1997, A\&A, 367, 59
\reference{} Pierre, M., Le Borgne, J.F., Soucail, G., \& Kneib, J.P. 1996,
A\&A, 311, 413P
\reference{} Rines, K., Geller, M.J., Diaferio, A., Mohr, J.J., \& Wegner, 
G.A. 2000, AJ, 120, 2338
\reference{} Smail, I., Ellis, R.S., Dressler, A., Couch, W., Oemler, A.,
Sharple, R.M., \& Butcher, H. 1997, ApJ, 479, 70
\reference{} Wu, X.-P., Xue, Y.-J., \& Fang, L.-Z. 1999, ApJ, 524, 22
\reference{} Wu, X.-P. 2000, MNRAS, 316, 299
\reference{} Viana, P.T.P. \& Liddle, A.R. 1999, MNRAS, 303, 535
\reference{} Yee, H.K.C., Ellingson, E.,  \& Carlberg, R.G. 1996, ApJS, 
102, 269 
\reference{} Yee, H.K.C., Ellingson, E., Morris, S.L., Abraham, R.G.,
 \& Carlberg, R.G. 1998, ApJS, 116, 211
\reference{} Yee, H.K.C. \& Gladders, M.D. 2001, in the proceedings
of {\it AMiBA 2001: High-z Clusters, Missing Baryons,
and CMB Polarization}, eds. L.-W. Chen et al.,  ASP Conf. Ser., 257, 201
\reference{} Yee, H.K.C. \& L\'opez-Cruz, O. 1999, AJ, 117, 1985 (YL99)
\reference{} Zwicky, F. 1937, ApJ, 86, 217

\end{references}
\end{document}